\documentclass[preprint,showpacs,preprintnumbers,amsmath,amssymb]{revtex4}
\usepackage[dvips]{graphicx}
\usepackage{rotating}
\usepackage{hyperref}
\usepackage{subfig}
\usepackage[normalem]{ulem}
\usepackage{color}
\usepackage{textcomp}
\definecolor{orange}{rgb}{1,0.5,0}
\usepackage[T1]{fontenc}

\begin{document}

\title{Majorana neutrinos production at LHeC in an effective approach}

\author{Luc\'{\i}a Duarte}
\email{lucia@fisica.edu.uy}
 \affiliation{Instituto de F\'{\i}sica, Facultad de Ingenier\'{\i}a,
 Universidad de la Rep\'ublica \\ Julio Herrera y Reissig 565,(11300) 
Montevideo, Uruguay.}

\author{Gabriel A. Gonz\'alez-Sprinberg}
 \affiliation{Instituto de F\'{\i}sica, Facultad de Ciencias,
Universidad de la Rep\'ublica \\ Igu\'a 4225,(11400) Montevideo, Uruguay.}

\author{Oscar A. Sampayo}
\email{sampayo@mdp.edu.ar}

 \affiliation{Instituto de Investigaciones F\'{\i}sicas de Mar del Plata 
(IFIMAR)\\ CONICET, UNMDP\\ Departamento de F\'{\i}sica,
Universidad Nacional de Mar del Plata \\
Funes 3350, (7600) Mar del Plata, Argentina.}

\begin{abstract}
We investigate the possibility of detecting Majorana neutrinos at the Large 
Hadron-electron Collider, an electron-proton collision mode at CERN.
We study the $l_j^{+} +  3 jets$ ($l_j\equiv e ,\mu ,\tau$)  final states that 
are, due to leptonic number violation, a clear signature for intermediate 
Majorana neutrino contributions. Such signals are not possible if the heavy 
neutrinos have Dirac nature. The interactions between Majorana neutrinos and 
the Standard Model particles are obtained from an effective Lagrangian 
approach. We present our results for the total cross section as a function of 
the neutrino mass, the effective couplings and the new physics scale. We also 
show the discovery region as a function of the Majorana neutrino mass and the 
effective couplings. Our results show that the LHeC may be able to discover 
Majorana neutrinos with masses lower than $700$ and $1300$ GeV for electron 
beams settings of $E_e=50$ GeV and $E_e=150$ GeV, respectively.
\end{abstract}


\maketitle

\section{\bf Introduction}

The discovery at the LHC of a new neutral boson has been a great scientific 
achievement for particle physics, and up to now no new physics has been found 
involving the electroweak scalar sector \cite{:2012gk, :2012gu}. Yet, it is 
well-known that the Standard Model (SM) -based on the gauge group $SU(3)_{c} 
\otimes SU(2)_{L} \otimes U(1)_{Y}$ undergoing a spontaneous symmetry breaking 
in its electroweak sector, up to the universal $U(1)_{EM}$- leaves important 
questions unexplained.
In the recent years, the first discovery of physics beyond minimum SM has taken 
place through the observation of flavor neutrino oscillations. This also has 
led to nonzero neutrino masses below or of the order of an electron volt 
\cite{Agashe:2014kda}. Considering this scenario, physics at new colliders 
should probe not only the mechanism behind electroweak symmetry breaking and 
the stabilization of the electroweak scale, but also trace the existence and 
nature of neutrino masses. The recent Large Hadron-electron Collider (LHeC) proposal \cite{Bruening:2013bga}, an 
electron-proton collider at CERN, could serve both purposes.

Neutrino masses are difficult to generate in a natural way in the SM Yukawa 
interactions framework, and a very attractive and well-known scheme to obtain 
them is the seesaw mechanism, which requires the presence of heavy right-handed 
neutrino species of the Majorana type that allow for lepton number violation 
(LNV) \cite{GellMann:1980vs, 
Yanagida:1980xy,Mohapatra:1979ia,Minkowski:1977sc}. The discovery of Majorana 
neutrinos would have profound theoretical implications in the formulation of a 
new model framework, while yielding insights into the origin of mass itself. 
Observation of any LNV process would be of great impact on particle physics and 
cosmology as, if neutrinos are Majorana particles, they may fit into the 
leptogenesis scenario for creating the baryon asymmetry, and hence the ordinary 
matter of the Universe \cite{Fukugita:1986hr}. However, the minimal seesaw 
framework generally leads to the decoupling of the Majorana neutrinos, and the 
observation of any LNV signal would indeed point toward new 
physics beyond the minimal seesaw models \cite{delAguila:2008ir}. 
In this work we will investigate the possibility of discovering Majorana 
neutrinos at the LHeC, considering its interactions in a general and 
model-independent effective Lagrangian approach.

The seesaw mechanism, among other SM extensions, requires one or more extra 
right-handed neutrinos $\nu_{R}$ with a 
mass term
\begin{eqnarray}
\label{lmass} \mathcal L^{mass}= - \frac12 \bar{\nu_R^c} \;  M \;
\nu_R - \bar L \; \widetilde{\phi} \; Y  \; \nu_R + h.c. \;~ ,
\end{eqnarray}
where $L$ denotes the left-handed lepton doublet, $Y$ denotes the Yukawa coupling 
matrix,
$\phi$ denotes the Higgs doublet and $M$ denotes the Majorana neutrino mass. 

The diagonalization of the mass term gives
\begin{eqnarray}
\label{mnu} m_{\nu}=m_{\mathcal D} M^{-1} m^T_{\mathcal D}, \;\;\;
\mbox{with} \;\;\; m_{\mathcal D} = Y \frac{v}{\sqrt{2}} \; ,
\end{eqnarray}
and a mixing angle $U_{lN} \sim m_{\mathcal D}/M$ between the light and the 
heavy Majorana neutrinos $N$ \cite{Kayser:1989iu,Mohapatra:1998rq}.

The mixing angle $U_{lN}$ weighs the coupling of $N$ with the SM particles and 
in particular with the charged leptons through the
$V-A$ interaction:
\begin{eqnarray}
\label{lw}
 \mathcal L_W = -\frac{g}{\sqrt{2}} U_{lN} \overline N^c 
\gamma^{\mu} P_L l W^+_{\mu} + h.c.
\end{eqnarray}

In typical seesaw scenarios, the Dirac mass terms are expected to be around the 
electroweak scale ($m_{\mathcal D}\sim m_{W}$) in order to have Yukawa 
couplings $Y\sim O(1)$ in Eq.(\ref{mnu}), whereas the Majorana mass $M$ -being 
a singlet under the SM gauge group- may be very large, close to the grand unification ccale. Thereby, the seesaw mechanism can explain the smallness of 
the observed light neutrino masses ($m_{\nu}\sim0.01$ eV) while leading to the 
decoupling of $N$. Even a different choice in which $M\sim100$ GeV and 
$m_{\mathcal D}\sim 0.1\; m_{e}$, keeping $m_{\nu}\sim 0.01$ eV, implies a 
vanishing mixing angle $U_{lN}\sim 10^{-7}$ \cite{delAguila:2008ir}. This 
effect is so weak that the observation of LNV  must indicate new physics beyond 
the minimal seesaw mechanism, as was indicated in Ref.\cite{delAguila:2008ir}.

In view of the above discussion, in this work, we consider -in a 
model-independent way- the effective interactions of the Majorana neutrino $N$ 
with a mass value lower than the new physics scale $\Lambda$ and a negligible 
mixing to $\nu_L$ . In the case that heavy neutrinos do exist, present and 
future
experiments will be capable of determining their nature. In particular, the 
production of Majorana neutrinos via $e^+e^-$, $e^- \gamma$, $\gamma
\gamma$ and hadronic collision have been extensively investigated in the past 
\cite{delAguila:2008ir,Ma:1989jpa,Datta:1993nm,Gluza:1994ac,Hofer:1996cs,
Cvetic:1998vg,Almeida:2000pz,
Peressutti:2001ms,Peressutti:2002nf,Peressutti:2011kx,PhysRevD.90.013003,
Belanger:1995nh,Atre:2009rg}.

In this paper we study the possibility for an $e^- p$ collider at CERN (LHeC) 
in order to produce clear signatures of Majorana neutrinos in
the context of interactions coming from an effective Lagrangian approach. We 
study the lepton number violating reaction $e^- p \rightarrow l_j^{+}  +
3 jets$ ($l_j\equiv e,\mu,\tau$) which receives contributions from the diagrams 
from the processes depicted in Fig.\ref{fig:euNldu}. We have not considered the pure lepton decay channels because they
involve light neutrinos that escape detection, in which case the Majorana 
nature of the heavy neutrinos would have no effect on the
signal, since we should be able to know whether the final state contains 
neutrinos or antineutrinos. 

The lepton number violating process studied here was previously investigated in 
Refs.\cite{Ingelman:1993ve, Buchmuller:1991tu}, for the type-I seesaw 
mechanism, focusing on the DESY experiment and extended to the LEP and LHC. Recent 
studies of the seesaw model at lepton-proton colliders like the LHeC were 
performed in Refs.\cite{Blaksley:2011ey, Liang:2010gm}. 

The principal advantage of electron-proton collisions with 
respect to hadron colliders is the cleanness of the signal. In the case of the 
LHeC, the leptonic number violation by $2$ units is ensured by the presence of 
a final antilepton. Conversely, lepton number violation detection in hadron 
colliders implies tagging two leptons of the same sign in the final state, 
together with a higher number of jets, making the signal more challenging to 
search for. 
In Ref. \cite{delAguila:2008ir} the process $pp\rightarrow l^{+} 
l^{+} jjjj$ is studied with the same effective formalism we apply here, and the 
authors claim that is possible to expect a 5$\sigma$ same-sign lepton signal 
for $m_N\le 600$GeV. As will be shown, we expect a significant signal for 
larger masses, in particular, for $E_e \le 50$GeV we expect $m_N \lesssim 
700$GeV, and for $E_e \le 150$GeV, we expect $m_N \lesssim 1300$GeV. The ATLAS collaboration has published new physics searches in the same-sign dilepton signal for this model \cite{Aad:2011vj, ATLAS:2012ak}, finding limits for the Majorana neutrino mass and certain effective couplings.

In Sec. \ref{sec:effL} we review the effective Lagrangian approach and 
present our results for the scattering amplitudes. The numerical results are 
presented in Sec. \ref{subsec:NumRes}, including the SM backgrounds, the 
neutrinoless double-$\beta$ decay bounds considered, and the obtained cross 
sections and discovery regions for the Majorana neutrino. Our conclusions are 
presented in Sec. \ref{sec:Concl}.

\section{\bf Effective Lagrangian and Scattering Amplitudes}\label{sec:effL}

The effects of new physics beyond the SM can be parametrized by a series of 
effective operators $\mathcal{O}$
constructed with the SM and the Majorana neutrino fields
and preserving the $SU(2)_L \otimes U(1)_Y$ gauge
symmetry \cite{Buchmuller:1985jz, Grzadkowski:2010es}. These effective 
operators represent the
low-energy limit of an unknown theory, and their effects are suppressed
by inverse powers of the new physics scale $\Lambda$. We consider the lowest-order new physics terms, taking into account only
dimension-$6$ operators and nonviolating baryon number interactions and 
discarding the operators generated at one-loop level in the underlying full 
theory, as they are naturally suppressed by a $\mathcal{O}\sim 1/{16\pi^2}$ 
factor \cite{Arzt:1994gp,delAguila:2008ir}.

The total Lagrangian is organized as:
\begin{eqnarray}
\mathcal{L}=\mathcal{L}_{SM}+\sum_{\mathcal{J},i}\frac{\alpha^{(i)}_{\mathcal{J}
}}{\Lambda^{2}}
\mathcal{O}_\mathcal{J}^{i}
\end{eqnarray}
where the indices $\mathcal{J}$ and $i$ label the operators and families
respectively. For the considered operators we follow 
Ref.\cite{delAguila:2008ir} starting with a rather general effective Lagrangian
density for the interaction of a Majorana neutrino $N$ with leptons
and quarks. All the operators listed here are generated at tree level in the 
unknown
fundamental high-energy theory.
The operators involving scalars and vectors are
\begin{eqnarray}
\mathcal{O}^i_{LN\phi}=(\phi^{\dag}\phi)(\bar L_i N \tilde{\phi}),
\;\; \mathcal{O}^i_{NN\phi}=i(\phi^{\dag}D_{\mu}\phi)(\overline N
\gamma^{\mu} N), \;\; \mathcal{O}^i_{Ne\phi}=i(\phi^T \epsilon
D_{\mu} \phi)(\overline N \gamma^{\mu} e_i)
\end{eqnarray}
and for the baryon-number conserving 4-fermion contact terms, we
have
\begin{eqnarray}
\mathcal{O}^i_{duNe}=(\bar d_i \gamma^{\mu} u_i)(\overline N \gamma_{\mu}
e_i) &,& \;\; \mathcal{O}^i_{fNN}=(\bar f_i \gamma^{\mu} f_i)(\overline N
\gamma_{\mu}
N), \\
\mathcal{O}^i_{LNLe}=(\bar L_i N)\epsilon (\bar L_i e_i)&,& \;\;
\mathcal{O}^i_{LNQd}=(\bar L_i N) \epsilon (\bar Q_i d_i), \\
\mathcal{O}^i_{QuNL}=(\bar Q_i u_i)(\overline N L_i)&,& \;\;
\mathcal{O}^i_{QNLd}=(\bar Q_i N)\epsilon (\bar L_i d_i),\\
\mathcal{O}^i_{LN}=|\bar{L}_i N|^2&&
\end{eqnarray}
where $e_i$, $u_i$, $d_i$ and $L_i$, $Q_i$ denote the $SU(2)$ right-handed 
singlets and left-handed doublets, respectively. These
are the contributing operators to the Majorana neutrino $N$ production and 
decay processes.

The relevant effective Lagrangian terms contributing to the production process 
considered are:
\begin{eqnarray}\label{leff}
\mathcal{L}^N_{eff}&=&\frac{1}{\Lambda^2}\left\{- \frac{m_W v}{\sqrt{2}} 
\alpha^{(i)}_W
\; W^{\dag\; \mu} \; \overline N_R \gamma_{\mu} e_{R,i} + \alpha^{(i)}_{V_0} 
\bar d_{R,i} \gamma^{\mu} u_{R,i} \overline N_R \gamma_{\mu}
e_{R,i} + \right.
\nonumber
\\ &&
 \alpha^{(i)}_{S_1}(\bar
u_{L,i}u_{R,i}\overline N \nu_{L,i}+\bar d_{L,i}u_{R,i} \overline N e_{L,i})
 +
\alpha^{(i)}_{S_2} (\bar \nu_{L,i}N_R \bar d_{L,i}d_{R,i}-\bar
e_{L,i}N_R \bar u_{L,i}d_{R,i}) +
\nonumber
\\ &&
\left. \alpha^{(i)}_{S_3}(\bar u_{L,i}N_R
\bar e_{L,i}d_{R,i}-\bar d_{L,i}N_R \bar \nu_{L,i}d_{R,i})
  + h.c. \right\}
\end{eqnarray}
where the sum over $i$ is understood  and the constants
$\alpha^{(i)}_{\mathcal{J}}$ are associated to specific operators
\begin{eqnarray}\label{alphas}
\alpha^{(i)}_W=\alpha^{(i)}_{Ne\Phi},\;
\alpha^{(i)}_{V_0}=\alpha^{(i)}_{duNe},\;\;
\alpha^{(i)}_{S_1}&=&\alpha^{(i)}_{QuNL},\;
\alpha^{(i)}_{S_2}=\alpha^{(i)}_{LNQd},\;\;
\alpha^{(i)}_{S_3}=\alpha^{(i)}_{QNLd}~.\;
\end{eqnarray}
\begin{figure*}[h]
\begin{center}
\includegraphics[totalheight=7cm]{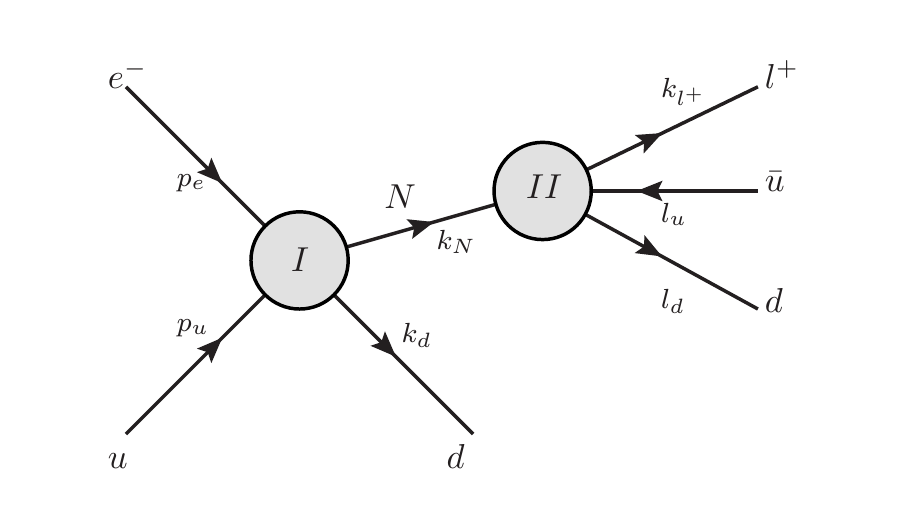}
\caption{\label{fig:euNldu}  Diagrams contributing to the production of 
Majorana neutrinos in $e p$ colliders.}
\end{center}
\end{figure*}
Using the effective Lagrangian in Eq.(\ref{leff}), we calculate the cross 
section for the production of the Majorana
neutrino according to the processes shown in Fig.\ref{fig:euNldu}.
Taking the center of mass energy $\sqrt{s}=\sqrt{4 E_e E_p}$, $\hat{\sigma}$ and $\hat{s}$ 
to be the parton level scattering cross section, and the squared center-of-mass energy, and with $x$ the usual deep inelastic scaling variable, 
we obtain
\begin{eqnarray}
\sigma(ep\rightarrow l^+ + 3 jets)=\sum_i \int_{m_N^2/s}^1 dx f_i(x) \hat 
\sigma_i(x s)
\end{eqnarray}
where $i=1$ corresponds to the channel $e u \rightarrow N d$ and $i=2$ corresponds to the 
crossed channel $e \bar d \rightarrow N \bar u$ obtained by the crossing 
symmetry. 
The function $f_1(x)$ represents the $u(x)$ parton distribution function (PDF), 
and $f_2(x)$ represents the one for $\bar d(x)$ and
\begin{eqnarray}
\hat \sigma_i(x s)= \int (2\pi)^4 
\delta^{(4)}(p_e+p_u-\sum_{j=1,4} 
k_j)\overline{|M_{(i)}|^2} \prod_{j=1,4}\frac{d^4k_j}{2\pi^3} ~.
\end{eqnarray}
The squared scattering amplitudes in the {\it narrow width aproximation } are
\begin{eqnarray}
\overline{\vert M_{(i)} \vert}^2=\left(\frac{\pi}{4m_N 
\Gamma_N \hat{s}}\right)\delta(k_N^2-m_N^2) \vert \Lambda_{(I),i} \vert^2 
(\vert \Lambda_{(II)}^{(+)} \vert^2 + \vert \Lambda_{II}^{(-)} \vert^2)
\end{eqnarray}
where
\begin{eqnarray}
\vert \Lambda_{(I),1} \vert^2 &&= \frac{4}{\Lambda^2}\left[ 
(\alpha_{S_{2}}(\alpha_{S_{2}}-\alpha_{S_{3}})+\alpha^2_{S_{1}}) (k_d \cdot 
p_u)(k_N \cdot p_e)+          
\right. \nonumber \\  &&\left. (4 \alpha_{W}^2 \vert \Pi^{(2)}_W \vert^2 
+\alpha_{S_{3}}(\alpha_{S_{3}}-\alpha_{S_{2}}))(k_d \cdot p_e)(k_N \cdot p_u)+ 
(\alpha_{S_{3}} \alpha_{S_{2}}+4 \alpha^ 2_{V_{0}}) (k_d \cdot k_N)(p_e \cdot 
p_u) \right]
\nonumber\\
\nonumber\\
\vert \Lambda_{(II)}^{(-)} \vert^2 &&= \frac{16}{\Lambda^4}\left[\vert 
\Pi^{(2)}_W \vert^2 \alpha_W^2 
(k_N \cdot l_u)(k_{l^+} \cdot l_d)+ \alpha_{V_0}^2 (k_N \cdot l_d)(k_{l^+} 
\cdot l_u) \right]
\nonumber\\
\nonumber\\
\vert \Lambda_{(II)}^{(+)} \vert^2 &&= 
\frac{4}{\Lambda^4}\left[(\alpha_{S_1}^2+\alpha_{S_2}^2-\alpha_{S_2}\alpha_{S_3}
)(l_u \cdot l_d)(k_{l^+} \cdot k_N) + \right.
\nonumber \\ &&\left. (\alpha_{S_3}^2-\alpha_{S_2}\alpha_{S_3})(k_{l^+} \cdot  
l_d)(l_u \cdot k_N)+
\alpha_{S_2} \alpha_{S_3} (l_u \cdot k_{l^+})(l_d \cdot k_N) \right]
\end{eqnarray}
 with $\Pi^{(1)}_W=m_W^2/(-2(p_u \cdot k_d)-m_W^2)$, $\Pi^{(2)}_W=m_W^2/(2(l_u 
\cdot l_d)-m_W^2)$. The final leptons can be either of $e^{+}$, $\mu^{+}$ or 
$\tau^{+}$ since this is allowed by the interaction Lagrangian 
(Eq.(\ref{leff})). All these possible final states are clear signals for 
intermediary Majorana neutrinos, and thus we sum the cross section over the flavors 
of the final leptons. The total width ($\Gamma_N$) for the Majorana neutrino decay is the 
calculated in Ref.\cite{Peressutti:2011kx}.

\section{\bf Numerical Results}\label{subsec:NumRes}

For the numerical study we assume an LHC-like beam of protons with an energy of 
$7$ TeV, while examining two choices for the electron beam. We consider a low-energy scenario with an electron beam of $E_e=50$ GeV (Scenario 1), and another 
high-energy scenario with $E_e=150$ GeV (Scenario 2). For each experimental 
setup we assume a baseline integrated luminosity of $L=100$ fb$^{-1}$ that is 
close to the values discussed for the LHeC proposal \cite{Bruening:2013bga}.

The branching ratios, cross sections and discovery regions for the Majorana 
neutrino in the effective Lagrangian approach considered in this paper depend 
on the quotient of the coupling constant $\alpha^{(i)}_{\mathcal{J}}$, 
associated with the operators in Eq.(\ref{leff}), and the new physics scale 
$\Lambda$ squared i.e. 
$\kappa^{(i)}_{\mathcal{J}}=\alpha^{(i)}_{\mathcal{J}}/\Lambda^{2}$, in 
addition to the Majorana neutrino mass $m_{N}$.
The considered operators are bounded by LEP and low-energy data and we have 
also taken into account the bounds on the operators that come from the 
neutrinoless double-$\beta$ decay ($0\nu_{\beta\beta}$-decay). 

We start this section discussing the SM backgrounds, the LEP, low-energy data 
and $0\nu_{\beta \beta}$-decay bounds, before showing our results for the 
scattering cross section for the process $e^- p \rightarrow l_j^{+} + 3 jets$, 
the different distributions and cuts implemented, and the Majorana neutrino 
discovery regions for both considered scenarios.

\subsection{\bf Standard Model background}\label{subsec:SMbckg}

The considered signal, being a lepton number violating process, is strictly 
forbidden in the Standard Model.
The SM background will always involve additional light neutrinos that escape 
the detectors and generate missing energy. This fact makes the signal very 
clean and difficult to mimic by SM processes. 

As was pointed out in Ref.\cite{Blaksley:2011ey}, the 
dominant background comes from $W$ production, with its subsequent decay into 
$l^+$ ($e^+,\mu^+,\tau^+$). In particular, the process $e^{-}p\rightarrow e^- 
l^+ j j j \nu$ is not distinguished from the signal if the outgoing electron is 
lost in the beam line. This process is dominated by the exchange of an almost 
real photon with a very collinear outgoing electron ($p\gamma\rightarrow l^+ j j 
j \nu$). This last process, convoluted with the PDF representing the 
probability of finding a photon inside an electron, is found to be the major 
contribution to $W$ production. The simulation of the background processes was 
done using the program CalcHep \cite{Belyaev:2012qa}. In Sect. \ref{subsec:cuts} we discuss different cuts to increase the sensitivity and 
improve the signal-to-background relation.
\subsection{LEP, low-energy, and neutrinoless double-$\beta$ decay 
bounds}\label{subsec:2beta}

The heavy Majorana neutrino couples to the three flavor 
families with couplings
 $\kappa^{(i)}_{\mathcal{J}}=\alpha^{(i)}_{\mathcal{J}}/\Lambda^{2}$. 
These couplings can be related with the mixing angle between light and heavy 
neutrinos $U_{lN}$, 
comparing the operator $\mathcal{O}^i_{Ne\phi}$ with the strength of the 
vector-axial vector interaction in Eq.\ref{lw}. 
The relation is $U_{l_{i}N}=\frac{v^2}{2}\frac{\alpha^{(i)}_{W}}{\Lambda^{2}}$ 
\cite{delAguila:2008ir}. 
The mixing angles $U_{lN}$ are bounded by LEP and low-energy data 
\cite{delAguila:2005pf, Bray:2005wv, Langacker:1988up, Nardi:1994iv, 
Bergmann:1998rg, Tommasini:1995ii}. 
In our case, with only one heavy neutrino $N$, and following the treatment made 
in Refs. \cite{delAguila:2005pf, Peressutti:2011kx}, 
we translate these model-independent bounds to the couplings $\kappa^{(i)}$, 
considering that all the operators satisfy the same and most stringent 
constraint given on Ref. \cite{Tommasini:1995ii} 
for $\Omega_{e \mu}=U_{eN}U_{\mu N}= \frac{v^2}{2}\kappa^{2}<1.0 \times 
10^{-4}$ with $v=250$ GeV. This leads to $\kappa<3.2\times10^{-7}$ GeV$^{-2}$, 
which, as we will show, is less restrictive than the the constraints imposed by 
the $0\nu_{\beta\beta}$-decay experiments.

To take into account the constraints imposed by the
$0\nu_{\beta\beta}$-decay experiments on some of the coupling
 constants $\alpha^{(i)}_{\mathcal J}$, we follow the developments presented in 
Refs.\cite{Mohapatra:1998ye,Rodejohann:2011mu}
and take the most stringent limits on the lifetime for neutrinoless
double-$\beta$ decay ($\tau_{{0\nu}_{\beta\beta}} \geq 2.1 \times
10^{25}$ yr) obtained by the GERDA collaboration \cite{Macolino:2013ifa}.

\begin{figure*}
\begin{center}
\includegraphics[totalheight=7cm]{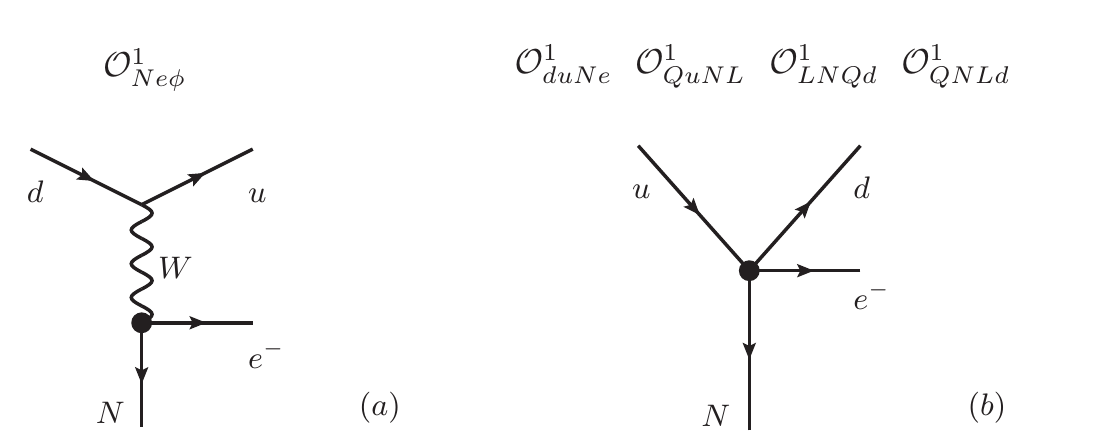}
\caption{\label{fig:2betaope} Contribution to $0\nu_{\beta \beta}$-decay. In 
the diagram (a), the solid dot represents the
operator ${\mathcal O}^1_{Ne\phi}$ and in the diagram (b) the dot represents the
4-fermion operators ${\mathcal O}^1_{duNe}$, ${\mathcal
O}^1_{QuNL}$,
 ${\mathcal O}^1_{LNQd}$ and ${\mathcal O}^1_{QNLd}$}.
\end{center}
\end{figure*}

The lowest-order contribution to $0\nu_{\beta \beta}$-decay from the
considered effective operators comes from those that involve the $W$
field and the 4-fermion operators with quarks $u$, $d$, the lepton
$e$ and the Majorana neutrino $N$:
\begin{eqnarray}
\label{2beope}
 \mathcal O^1_{N e \phi} \;,\; \mathcal O^1_{duNe} \;,\;
\mathcal O^1_{QuNL} \;,\; \mathcal O^1_{LNQd} \;,\; \mathcal
O^1_{QNLd}~.
\end{eqnarray}

The contribution of these operators to $0\nu_{\beta \beta}$-decay is shown in 
Fig.\ref{fig:2betaope}.

For the coupling constant associated with each operator we use the
generic name $\alpha_{0\nu_{}\beta\beta}$; that is to say
\begin{eqnarray}
\alpha_{0\nu\beta\beta}=\alpha^{(1)}_{Ne\phi}=\alpha^{(1)}_{duNe}=\alpha^{(1)}_{
QuNL}=\alpha^{(1)}_{LNQd}=\alpha^{(1)}_{QNLd}~.
\end{eqnarray}

To estimate the bounds on the different $\alpha_{\mathcal J}^{(i)}$ we 
consider the case in which all coupling constants $\alpha$ are nonzero with 
equal values, and the individual contributions of each operator are considered 
to act alone. The maximum value for the $\alpha$'s is limited by the 
$0\nu_{}\beta\beta$ bound.

Following the treatment made in Ref.\cite{Peressutti:2011kx}, we obtain the 
bound value for the quotient $\kappa_{0\nu \beta \beta}=\alpha_{0\nu \beta 
\beta}/\Lambda^{2}:$
\begin{equation}
\label{0n2bbound}
\kappa_{0\nu \beta \beta}=\frac{\alpha_{0\nu \beta \beta}}{\Lambda^{2}} \leq 
7.8 \times 10^{-8}
\left(\frac{m_N}{100GeV}\right)^{1/2}
\end{equation}

\subsection{Signal cross section}\label{subsec:CrossSect}

We have already discussed in the previous section that some of the operators 
that contribute
to the neutrinoless double beta decay ($0\nu_{\beta\beta}$-decay) may be 
strongly constrained. 
Therefore, for studying the Majorana neutrino production cross section in $e p$ 
colliders 
and the following decay $N \rightarrow l^+ + 2 jets$ we analyze two situations: 
in {\bf Set I} we consider the case in which the effective couplings
for the operators that do not contribute to neutrinoless decay take all the 
same value $\alpha=1$, and in {\bf Set II} we consider all those effective couplings to be equal and 
limited by the neutrinoless double beta decay bound Eq.(\ref{0n2bbound}).

The Majorana neutrino width
was studied in detail in Ref.\cite{Peressutti:2011kx}, in which all possible 
effective operators
of dimension-$6$ involving quarks were taken into account.

\begin{figure*}[h]
\centering
\subfloat[Signal cross section.]{\label{fig:sigma}\includegraphics[totalheight=5.8cm]{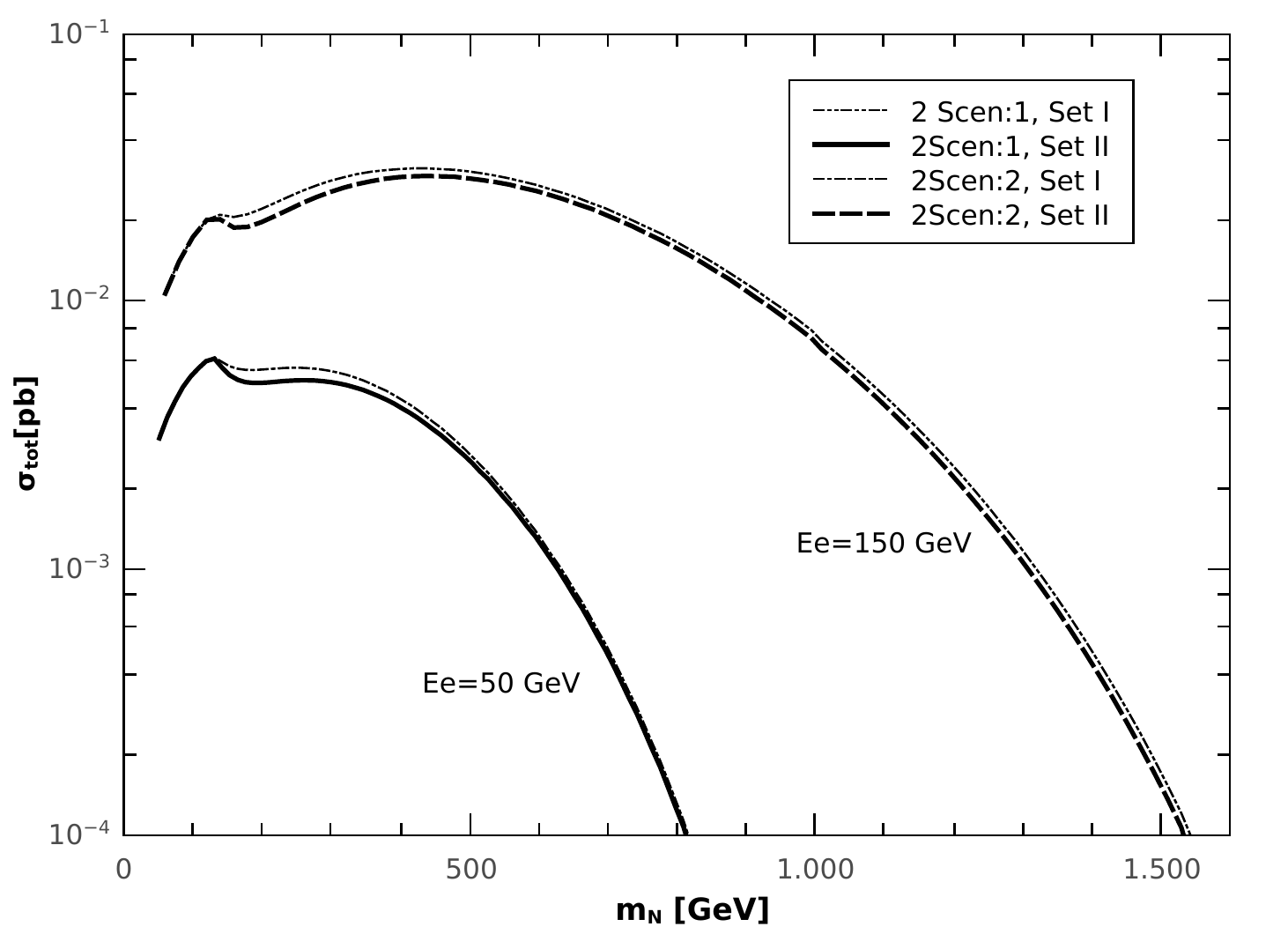}}~
\subfloat[Background $E_T$ dependence.]{\label{fig:cutEt}\includegraphics[totalheight=5.8cm]{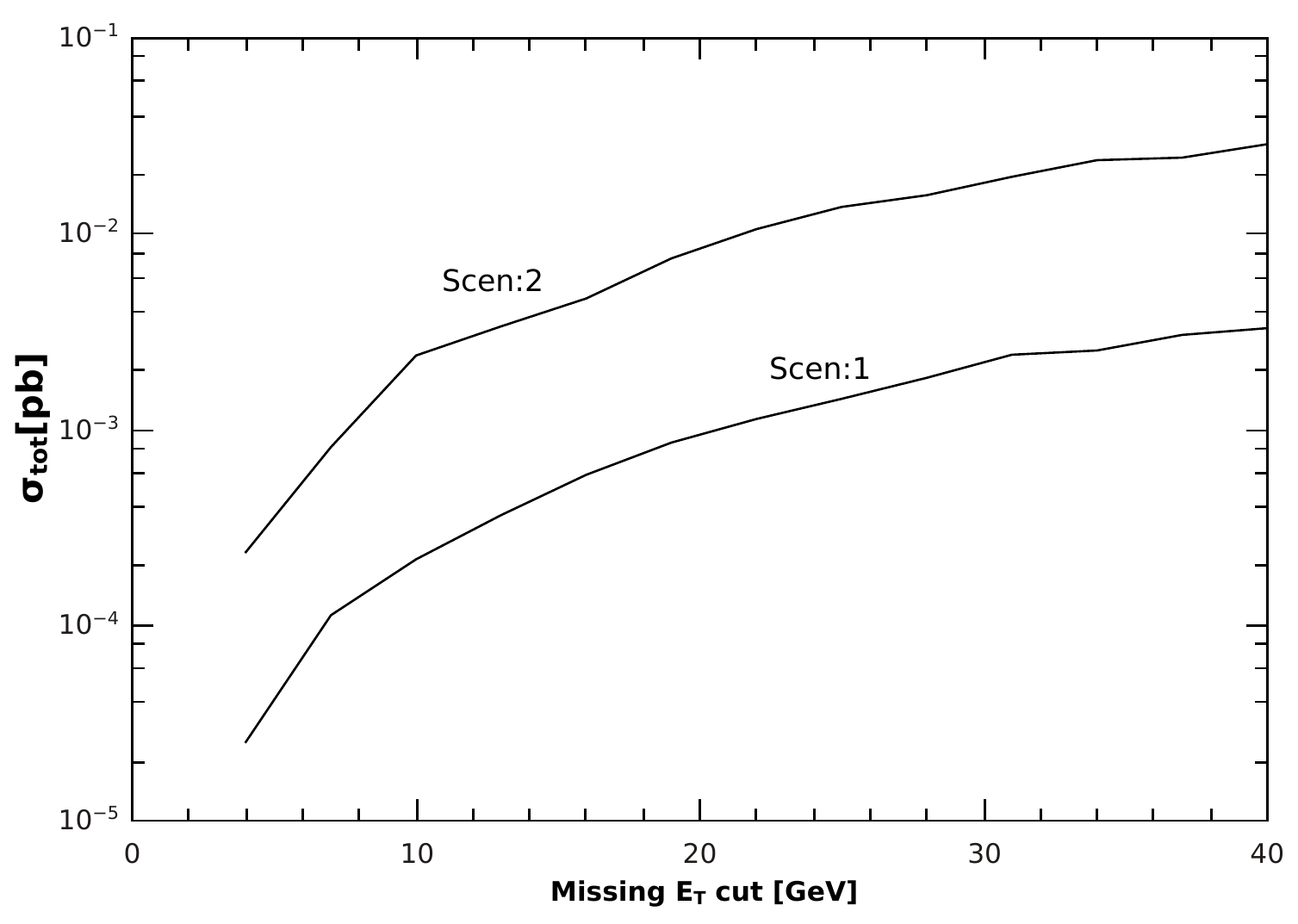}}
\caption{Cross section for the process $e p \rightarrow N X$ with $N$ decaying according to Ref.\cite{Peressutti:2011kx} (a) and background dependence with missing $E_T$ (b).}
\end{figure*}

In Fig.\ref{fig:sigma} we show the results for the cross
section, as a function of the
Majorana neutrino mass $m_N$, for the considered electron beam energies: 
$E_{e}=50$ 
GeV (Scenario 1) and $E_{e}=150$ GeV (Scenario 2) for both Sets {\bf I} and 
{\bf II}. The results are very similar for both sets.
We have considered $\sqrt{s} < \Lambda$ in order to ensure the validity of the 
effective Lagrangian approach. 
We display here the results for $\Lambda=2500$ GeV. 


The phase-space integration of the squared amplitude is made 
generating the final momenta with the Monte Carlo routine RAMBO 
\cite{Kleiss:1985gy}. This allows us to make the distributions and necessary 
cuts in the phase space to study the possibility of discovering Majorana 
neutrino effects.

\subsection{Distributions and kinematical cuts.}\label{subsec:cuts}
\begin{figure*}
\centering
\subfloat[Scenario 1]{\label{dptcutEt_c1}\includegraphics[totalheight=5.8cm]{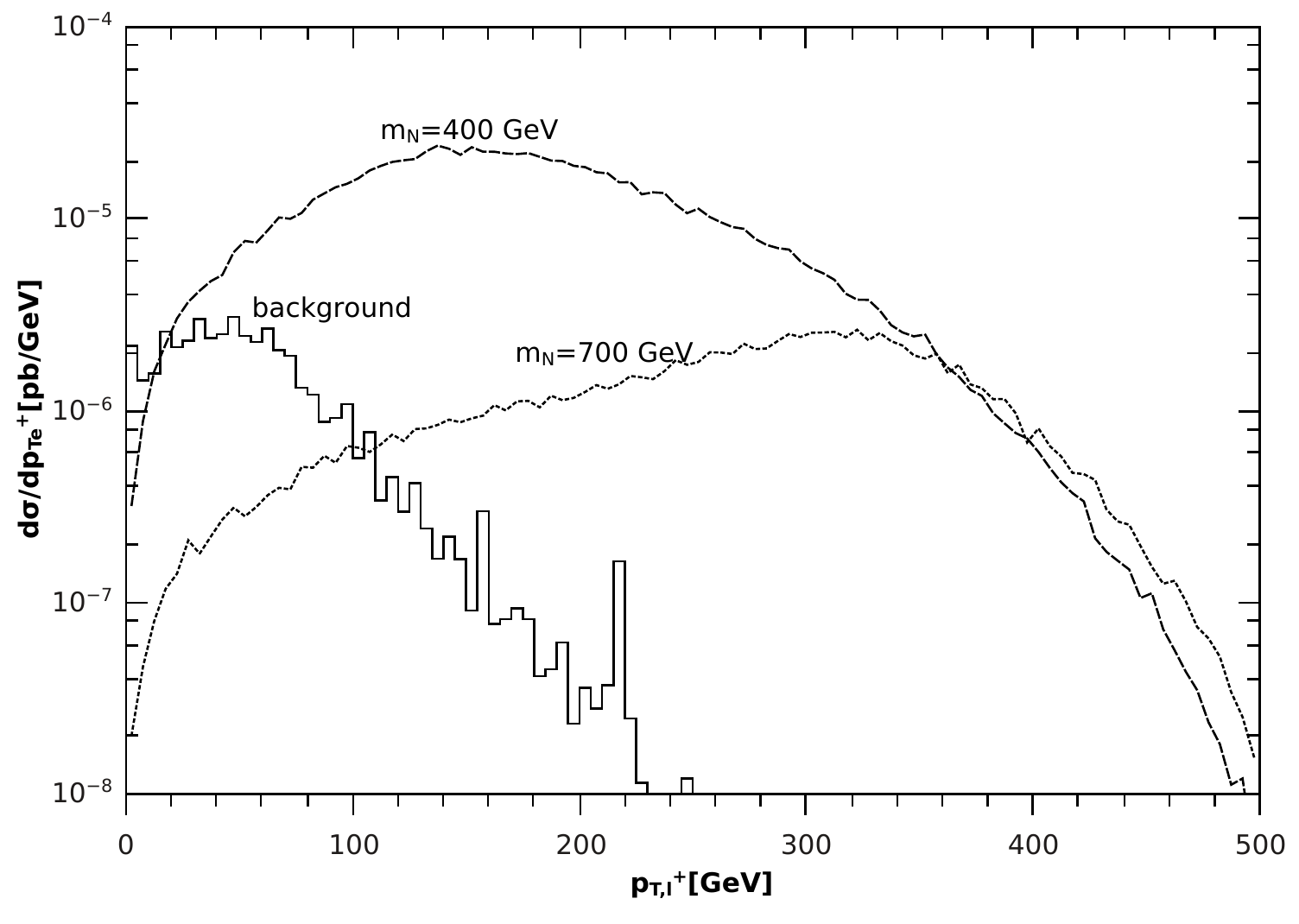}}
\subfloat[Scenario 2]{\label{dptcutEt_c2}\includegraphics[totalheight=5.8cm]{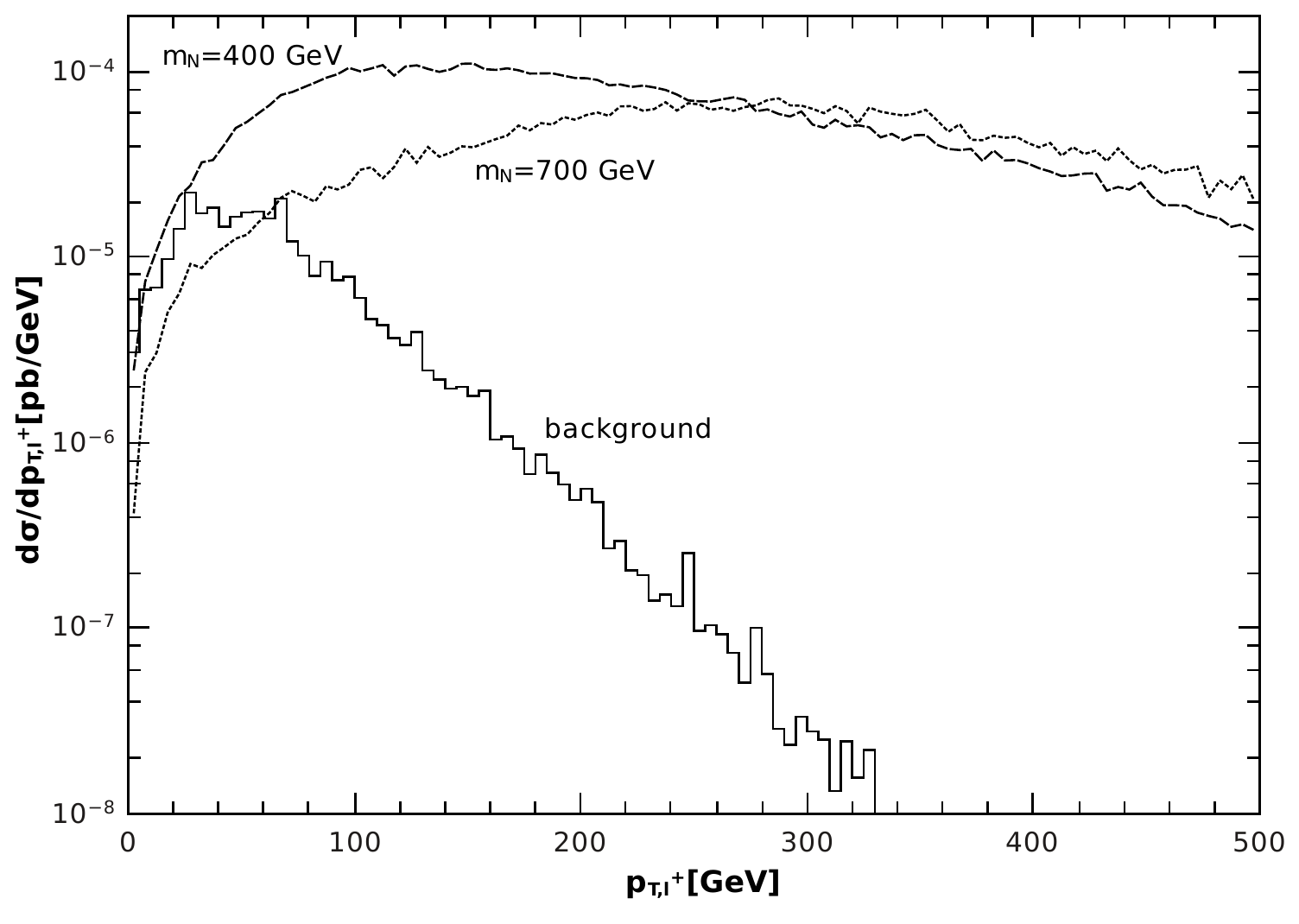}}
\caption{\label{fig:dptcutet} Differential cross section of signal and 
background in function of transverse momentum $p_{T,l^+}$. The cut in missing $E_{T}$ 
is included.}
\end{figure*}

\begin{figure*}
 \centering
\subfloat[Scenario 1]{\label{sig_bck_c1}\includegraphics[totalheight=5.58cm]{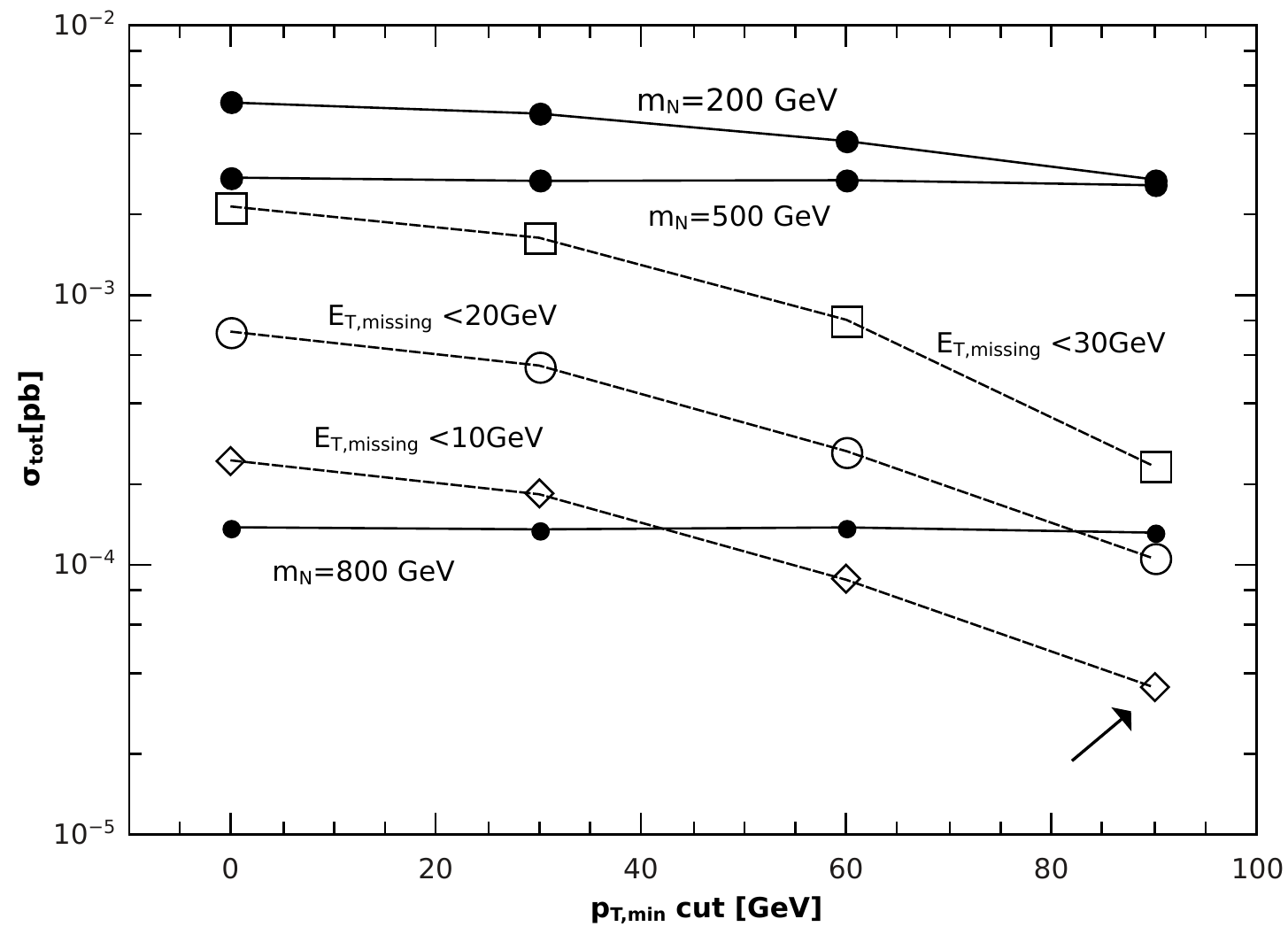}}
\subfloat[Scenario 2]{\label{sig_bck_c2}\includegraphics[totalheight=5.58cm]{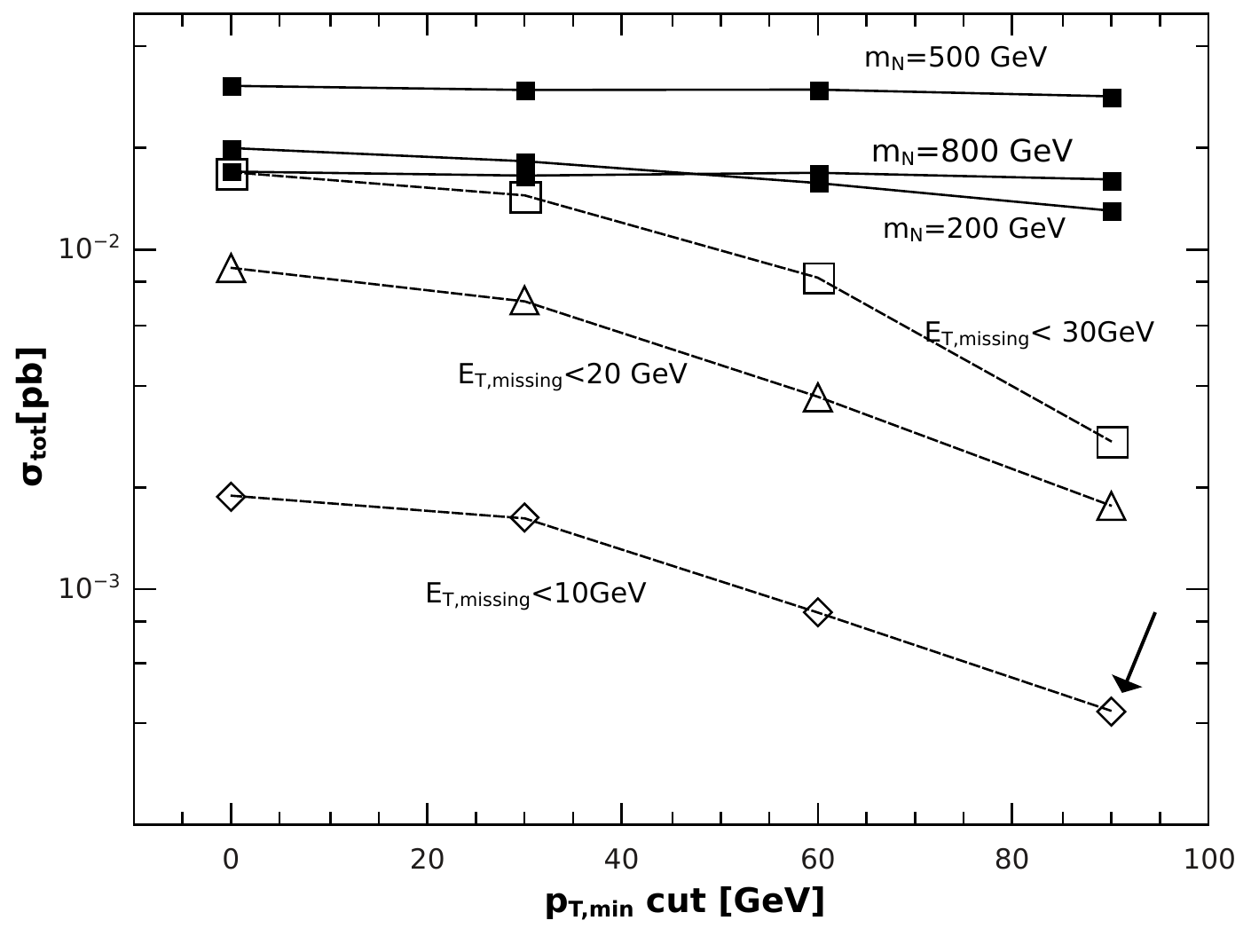}}
\caption{\label{fig:sigbck} Comparison between signal and background for 
different Majorana neutrino masses,  cut in missing $E_T$ and the transversal 
momentum of the final lepton $p_{T,l^ +}$. The solid lines show the cross 
section for the signal, and the dotted lines, show the cross section for the background. The arrows 
indicate the cuts and backgrounds used in the analysis.}
\end{figure*}

The dominant backgrounds for the studied process have been analyzed in Ref.
\cite{Blaksley:2011ey}. In particular, the authors conclude that a cut that could be 
effective to separate the signal and background is to reject events in which the 
outgoing $l^ +$ does not have a minimum transverse momentum. On the other hand, 
as the signal only includes visible particles and the background includes at 
least one neutrino, another possible cut is imposing an upper bound on 
the missing transverse energy. We follow this approach and implement the 
mentioned cuts. In Fig.\ref{fig:cutEt} we show the behavior of the background 
with the maximum missing energy $E_T$ for the scenarios in which $E_e=50$GeV 
(Scenario 1) and $E_e=150$GeV (Scenario 2). A cut of $E_{T,max}\le 10$ GeV, 
which is a reasonable value for the detector resolution, does not have appreciable 
effects on the signal but reduces the background significatively. 
In Fig.\ref{fig:dptcutet} we show the differential cross section for the 
background and the signal for different values of the Majorana mass as a 
function of the transverse momentum $p_{T,l^+}$ of the antilepton. In these figures the cut on the 
missing energy $E_T$ has already been included. As it can be appreciated, the 
background is mostly concentrated at low values of $p_{T,l^+}$, and a cut imposed on 
$p_{T,l^+}^{min}$ could be effective to improve the signal/background relation.
Finally, in Fig.\ref{fig:sigbck} we show a plot comparing the magnitude of the signal for different values of the Majorana neutrino mass (solid lines), and the background for different $E_{T,max}$ cuts (dashed lines), depending on the $p_{T,l^+}^{min}$ cut imposed. In both figures the arrows indicate the value of the cuts used in the analysis: we impose $p_{T,l^+} \ge 90$ GeV and $E_{T,miss} \le 10$ GeV in order to reduce the background without appreciably decreasing the signal.

%

%


\subsection{Discovery regions}\label{subsec:DiscReg}

To investigate the possibility of the detection of 
Majorana neutrinos in the process under consideration, we study the region
(discovery region) where the signal can be separated from the background with a 
statistical significance higher than $5\sigma$. We use the method of the {\it 
effective significance} described in Refs. \cite{Bityukov:2000tt, Narsky:1999kt}. 
There they show that the effective significance is well approximated by 
\begin{eqnarray}
\label{sensitivity}
\mathcal{S}=2(\sqrt{n_s+n_b}-\sqrt{n_b})-k(\alpha)
\end{eqnarray}
with k($\alpha$)=1.28 for $\alpha=0.1$ where $1-\alpha$ is the probability of measuring
a number of events bigger than a value $n_0$, such that the probability 
($\beta$) that the Standard Model reproduces such number is rather small, $\beta 
< 3 \times 10^{-7}$ for $\mathcal{S}>5$ ($5\sigma$ test).
In Eq.(\ref{sensitivity}) $n_s=L \sigma_s$ and $n_b=L \sigma_b$ are the numbers of 
events for the signal and backgrounds, with $L$ being the luminosity.

\begin{figure*}[h]
\begin{center}
\includegraphics[width=0.7\textwidth]{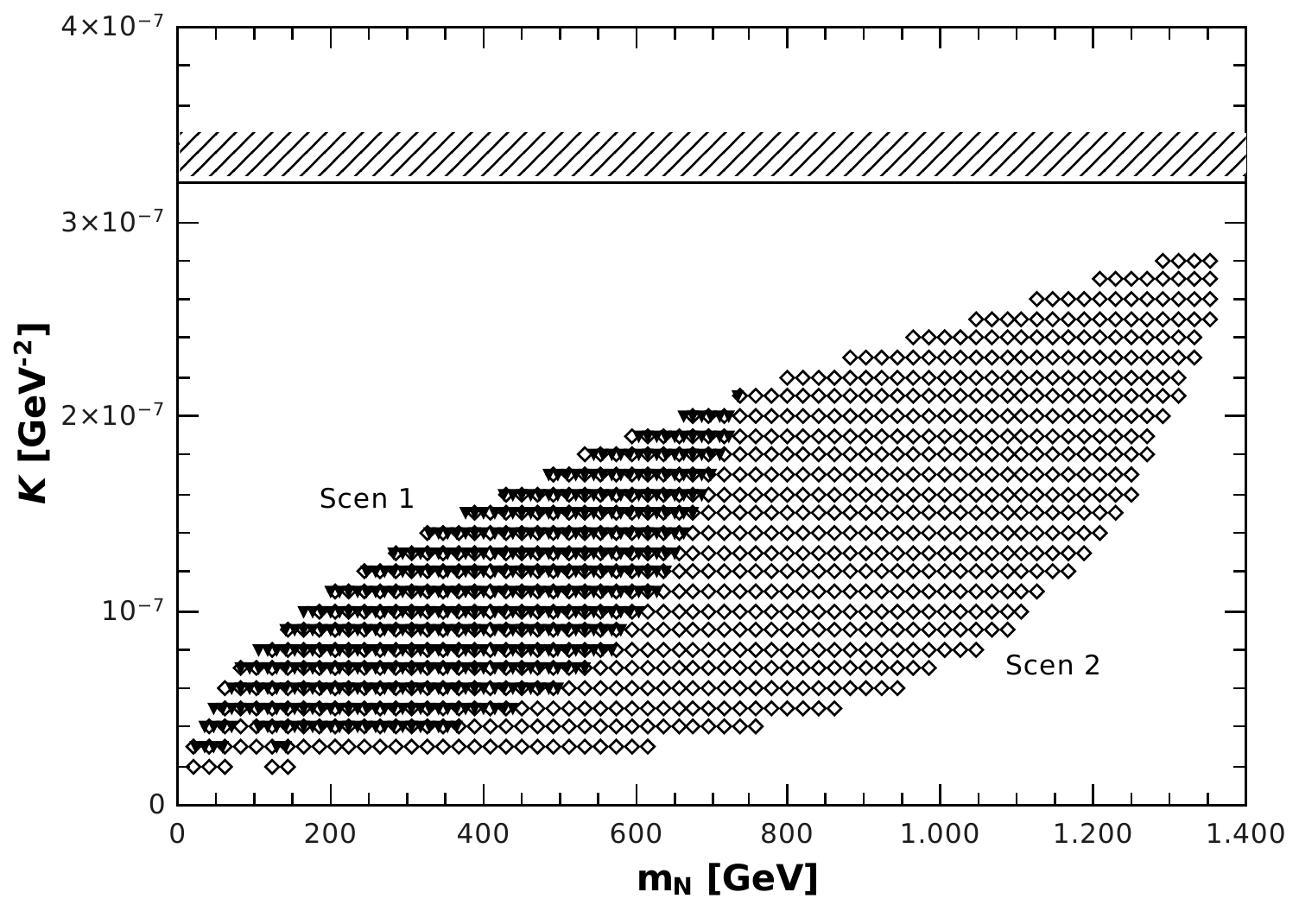}
\caption{\label{fig:discovery} Majorana neutrino discovery regions at $5 \, 
\sigma$. The horizontal line represents the low-energy and LEP limits discussed 
in Sec. \ref{subsec:2beta}.}
\end{center}
\end{figure*}

In Fig.\ref{fig:discovery} we show the discovery regions for different values 
of the Majorana neutrino mass $m_{N}$, and the quotient 
$\kappa^{(i)}_{\mathcal{J}}=\alpha^{(i)}_{\mathcal{J}}/\Lambda^2$. As we 
explained in Sec. \ref{subsec:2beta}, we consider the case in which all the 
$0\nu \beta \beta$ contributing coupling constants $\alpha^{(i)}_{\mathcal{J}}$ 
(generically $\alpha$) are nonzero and equal, so that $\kappa \leq 
\kappa_{0\nu \beta \beta}$ in Eq.\ref{0n2bbound}. The figure shows that 
Majorana neutrinos of masses up to near $1300$ GeV for Scenario 1, and $700$ 
GeV for Scenario 2 may be detected.

\begin{figure*}[h]
\begin{center}
\subfloat[Scenario 1]{\includegraphics[totalheight=5.8cm]{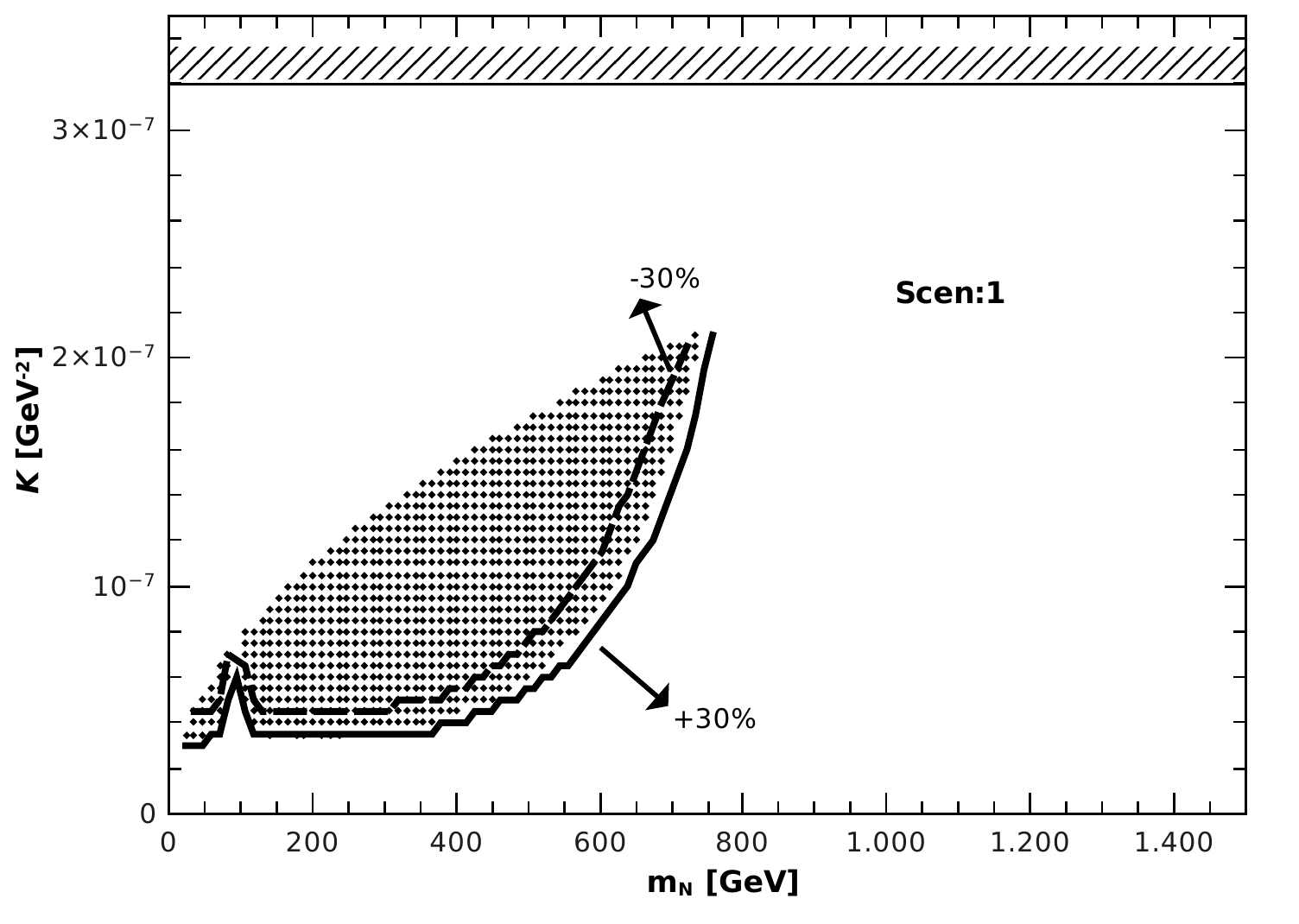}}
\subfloat[Scenario 2]{\includegraphics[totalheight=5.8cm]{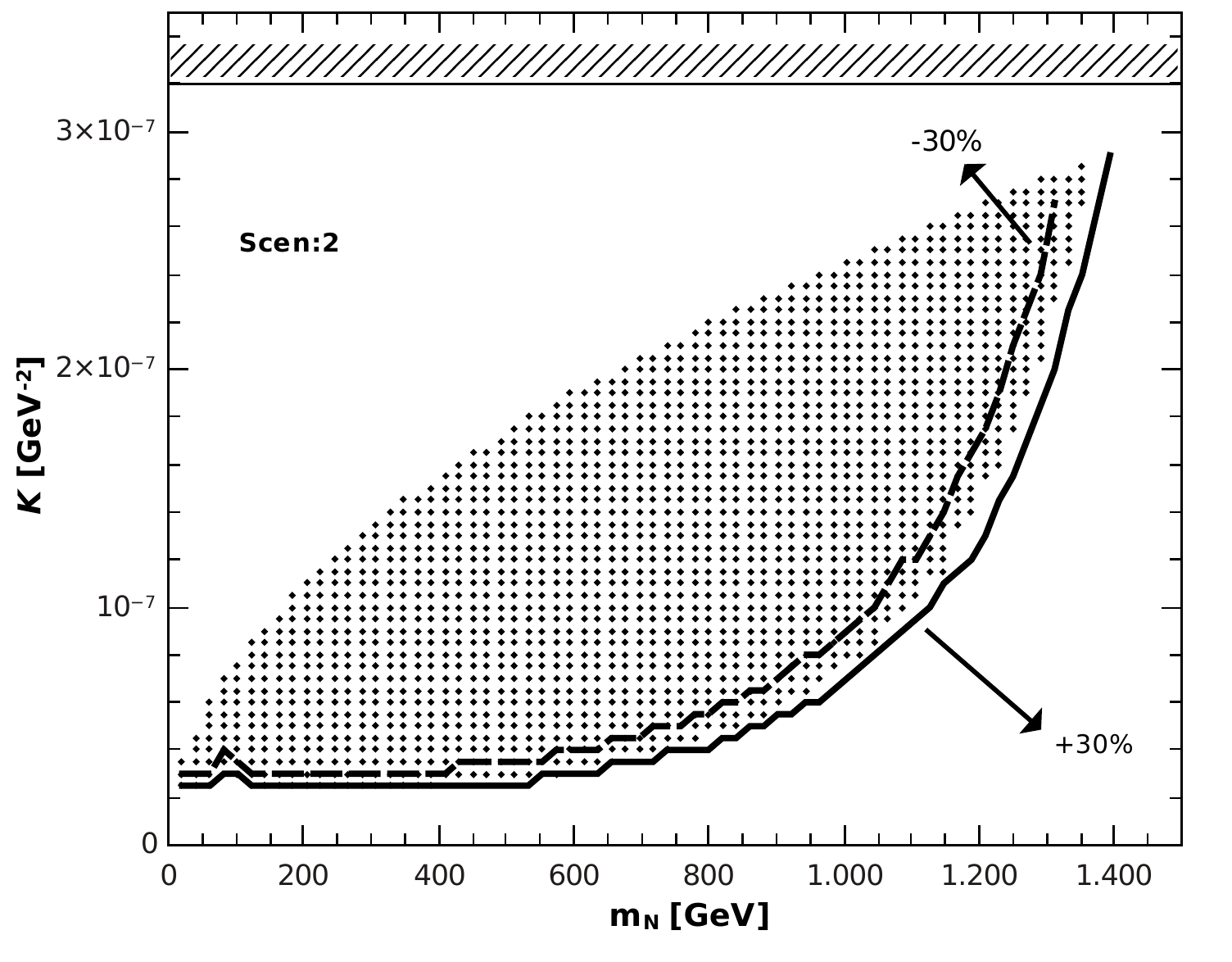}}
\caption{\label{fig:discovery12} Majorana neutrino discovery regions at $5 \, 
\sigma$, including systematic uncertainties in the signal.}
\end{center}
\end{figure*}


The maximum allowed value for the Majorana neutrino mass corresponds to the 
intersection 
between the $0\nu \beta \beta$ bound Eq.(\ref{0n2bbound}) and the contour of 
level $5$ for the surface $\mathcal{S}$ Eq.(\ref{sensitivity});
this is: $\mathcal{S}(m_N,\Lambda)=5$.  
The last equation can be written as $\alpha_{0\nu \beta\beta}/\Lambda^2\approx f(m_N)$ 
where $f$ is a function of $m_N$ and the collider energy 
but independent of $\Lambda$. Thus, the intersection and then the maximum 
possible value for $m_N$ is almost independent of the new physics scale 
$\Lambda$.

Systematic uncertainties are hard to estimate without a 
detailed reconstruction of the
detector, but they are expected to be around a few percent 
\cite{AbelleiraFernandez:2012cc}. However, the influence of the
background systematic uncertainties in the result is small because the 
background itself is small. In the case of the signal we have calculated
the modifications for the discovery region if the number of events for the 
signal is changed by $\pm30$\%. The results are shown in 
Fig.\ref{fig:discovery12}, showing no appreciable change in the region.

\section{Summary and conclusions}\label{sec:Concl}

To investigate the possibilities for discovering
Majorana neutrinos in an $e^- p$ collider at CERN (LHeC), we have calculated 
the cross section for the lepton number violating process
$e^- p \rightarrow l_j^{+} + 3 jets $ in an effective Lagrangian approach, 
complementing previous analyses for this facility involving typical seesaw 
scenarios.

The effective Lagrangian framework parameterizes new physics effects in a 
model independent way, allowing for sizable lepton number violating effects 
for effective couplings $\alpha^{(i)}_{\mathcal J}$ of order $1$, in contrast 
to the minimal seesaw mechanism, that leads to the decoupling of the Majorana 
neutrinos. 

While models like the minimal seesaw mechanism lead to the decoupling of the 
heavy Majorana neutrinos, predicting unobservable LNV, the effective Lagrangian 
framework considered in this work parameterizes the new physics effects in a 
model-independent way, enabling the occurrence of sizable LNV signals for 
effective couplings $\alpha^{(i)}_{\mathcal J}$ of order $1$. 

We have calculated the total unpolarized cross section $\sigma(e^- p 
\rightarrow l_j^{+} + 3 jets)$ for different values of $m_N$, the effective couplings 
$\alpha^{(i)}_{\mathcal J}$ and the new physics scale $\Lambda$, and implemented cuts in the phase space that can help to enhance the signal-to-background relation. 
We obtained the Majorana neutrino discovery regions at 5$\sigma$ statistical 
significance, combining the effect of the SM backgrounds with the most 
restrictive $0\nu_{\beta\beta}$-decay bounds for the effective couplings.

Our analysis shows that the LHeC facility could discover Majorana neutrinos 
with masses lower than $700$ and $1300$ GeV with a $7$ TeV proton beam, and 
electron beams of $E_e=50$ and $E_e=150$ GeV respectively. Thus, we find 
lepton-proton colliders could provide a new probe of the Majorana nature of 
neutrinos, shedding light on this fundamental unsolved issue in particle 
physics.

\

{\bf Acknowledgements}

We thank CONICET (Argentina) and Universidad Nacional de Mar del
Plata (Argentina); and PEDECIBA, ANII, and CSIC-UdelaR (Uruguay) for their 
financial support.

\bibliographystyle{bibstyle.bst}
\bibliography{Bib_N}

\end{document}